\documentclass[%
 reprint,
nofootinbib,
 amsmath,amssymb,
 aps,
]{revtex4-2}
\usepackage{amsmath}
\usepackage[colorlinks=true, linkcolor=blue, citecolor=blue]{hyperref}
\usepackage{graphicx}
\usepackage{dcolumn}
\usepackage{bm}

\begin{document}

\preprint{APS/123-QED}

\title{Generalized Virtual-Wave Theory for Photothermal Coherence Tomography under Arbitrary Excitation Toward Non-Contact Industrial Inspection of Composite Materials}

\author{Pengfei Zhu}
 \email{pengfei.zhu@bam.de}
\affiliation{%
 Bundesanstalt für Materialforschung and -prüfung (BAM), 12205 Berlin, Germany
}%

\author{Julien Lecompagnon}
\affiliation{%
	Bundesanstalt für Materialforschung and -prüfung (BAM), 12205 Berlin, Germany
}%

\author{Philipp Daniel Hirsch}
\affiliation{%
	Bundesanstalt für Materialforschung and -prüfung (BAM), 12205 Berlin, Germany
}%

\author{Mathias Ziegler}
\affiliation{%
	Bundesanstalt für Materialforschung and -prüfung (BAM), 12205 Berlin, Germany
}%

\date{\today}

\begin{abstract}
Photothermal imaging is a powerful noncontact and nondestructive technique for subsurface inspection of composite materials, yet its performance is fundamentally limited by the diffusive and irreversible nature of heat transport, leading to severe image blurring and ambiguous depth interpretation. The concept of virtual waves provides a route to overcome this limitation by linking diffusion fields to propagating wave fields, but existing approaches are largely restricted to idealized impulsive excitation. Here, we propose a generalized virtual-wave photothermal tomography framework that extends the diffusion-to-wave transformation to arbitrary boundary excitations, including pulsed, harmonic, and chirped waveforms. Starting from the heat equation with a general source term, we derive a Fredholm integral mapping between the measured diffusion field and a virtual wave field governed by a wave equation, explicitly enforcing causality and thermodynamic irreversibility. The resulting ill-posed inverse problem is solved using ADMM or truncated SVD, depending on the excitation characteristics. Numerical and experimental results demonstrate that the proposed method converts blurred thermal responses into wave-like fields with clear wavefronts and reflections, enabling improved depth localization and tomographic reconstruction. Experiments on carbon fiber reinforced polymer samples with embedded defects show enhanced contrast, sharper boundaries, and more reliable depth interpretation compared with conventional thermographic techniques. This work establishes a unified and physically grounded framework for wave-based photothermal tomography under realistic excitation conditions.
\end{abstract}

\maketitle

\section{Introduction}

Composite materials have been widely used in aerospace, energy, and advanced manufacturing due to their superior mechanical properties, lightweight characteristics, and corrosion resistance. However, their heterogeneous and multilayered structures make them particularly susceptible to hidden defects such as delamination, voids, and micro-cracks, which may severely compromise structural integrity and reliability. Therefore, developing effective non-destructive testing (NDT) techniques for composite materials is of critical importance for ensuring safety and performance in industrial applications. Photothermal imaging techniques were considered as a contactless and non-invasive method based on the thermal effect generated when materials absorb light radiation. It acquires information by measuring changes in thermal distribution on the surface or within materials caused by light excitation. Photothermal imaging techniques have been widely employed in chemical~\cite{ref1}, industrial~\cite{ref2}, biomedical~\cite{ref3}, and cultural heritage~\cite{ref4} fields. However, different from acoustic or electromagnetic methods, photothermal techniques are governed by the heat diffusion equation, which is a parabolic-type partial differential equation (PDE). This leads to the irreversible nature of photothermal techniques. Furthermore, the absence of wavefront and mathematically smoothing effect in thermal waves cause the image blurring and lack important high-frequency information.

Quantitative inversion based on thermal diffusion-wave field has been studied from the last century. Maldague et al.~\cite{ref5,ref6,ref7} proposed the phase-based method for evaluating the defect depth. The core idea is to solve the one-dimensional heat conduction equation with lock-in excitations. That will offer a quantitative relationship between depth and amplitude. Subsequently, they discovered that the depth information provided by the phase was \~1.78 times that of the amplitude~\cite{ref8,ref9,ref10}. Parker et al.~\cite{ref11} first derived the analytical solution for one-dimensional heat conduction equation with flash Dirac pulse excitation. Based on this analytical solution, one can define a thermal diffusion time $t_{\mathrm{diff}}=L^2/\alpha$ and a thermal diffusion length $L_{\mathrm{diff}} = \sqrt{\alpha t}$, which describe the typical scales over which heat spreads in time t, where $\alpha$ denotes the thermal diffusivity. Furthermore, Mandelis et al.~\cite{ref12} defines a quantitative relationship between time and depth based on the heat mass location. To this end, they proposed a match filtering-based photothermal coherence tomography technique~\cite{ref13}. This photothermal coherence tomography requires chirp-pulsed excitation and high framerate infrared camera. Subsequently, Zhu et al.~\cite{ref14} developed a frequency multiplexed photothermal correlation tomography technique for extending the chirp-pulsed excitation to arbitrary excitation. From the development of quantitative analysis, one can find that the one-dimensional quantitative detection is not sufficient for the increasing demands. Three-dimensional tomography has been obtaining more attention.

Image deblurring and denoising are also a hot issue in photothermal imaging fields. Pilla et al.~\cite{ref15,ref16} proposed a difference absolute contrast method, which is a simple but effective method for removing the background noise. The pulsed phase thermography (PPT) was then proposed by Maldague~\cite{ref17} based on the Fourier transform, which has been widely in real industrial inspection. Shepard et al.~\cite{ref18} proposed the thermal signal reconstruction (TSR) method, which is based on the mathematical relation between thermal signal and time in the logarithmic field. In addition, Rajic~\cite{ref19} proposed a principal component analysis-based method for three-dimensional thermal sequence data. The above signal processing techniques are famous in the infrared thermography fields such that every new developed method would like to compare with them. However, there is no obvious proof shown that which method is better than the conventional method with robustness and applicability. In particular, the artificial intelligence fails in this problem neither in supervised~\cite{ref20} nor unsupervised learning~\cite{ref21}. Recently, researchers tried to use structural illumination techniques~\cite{ref22,ref23,ref24} for achieving super-resolution imaging in photothermal fields. However, since photothermal imaging requires to acquire time-series data and the structural illumination techniques rely on multiple capturing with different patterns, the entire processing is extremely time-consuming compared with existed photothermal imaging techniques.

The above quantitative analysis and image denoising do not address the root cause of the problem. For instance, since the absence of wavefront, artificial definition of thermal diffusion length provides an ambiguous boundary resulting in inevitable errors when material or equipment changes. As for the image processing methods, they mostly repair the signal after diffusion degradation, rather than altering or constraining the physical processes themselves that cause the degradation. Recently, Burgholzer et al.~\cite{ref25,ref26,ref27} proposed the concept of virtual waves, which links the diffusion equation to the wave equation under Dirac pulse excitation. In this case, the photothermal imaging results will be altered to the ultrasound results. This provides a new mind for photothermal signal reconstruction~\cite{ref28,ref29,ref30}.

In this work, we use a more reasonable theoretical derivation in mathematics – considering a general boundary-injected forms of heat excitation rather than considering the Dirac pulse as the initial condition. Under this formalism, we can extend the virtual wave method from the Dirac pulse excitation to arbitrary excitation, which allows us to use more advanced signal processing techniques in radar, lock-in, or coded signal analysis. The theory of the generalized virtual wave photothermal tomography technique was derived. And numerical simulation and photothermal experiments are employed to validate the feasibility and accuracy of this proposed method.

\section{Generalized virtual wave photothermal tomography technique}
\subsection{Principle}
The heat conduction in isotropic medium is governed by the heat diffusion equation~\cite{ref31}:
\begin{equation}
	\left(\nabla^2 - \frac{1}{\alpha}\frac{\partial}{\partial t}\right)
	T(\mathbf{r}, t)
	= -\frac{1}{\kappa} Q(\mathbf{r}, t),
	\label{eq1}
\end{equation}
where $T(\mathbf{r}, t)$ denotes the temperature field, $\alpha$ denotes the thermal diffusivity, $\kappa$ denotes the thermal conductivity, and $Q(\mathbf{r}, t)$ denotes an externally applied heat source. Equation~\eqref{eq1} is obtained by rearranging the standard heat equation into a Helmholtz-like operator form for later comparison.

A scalar wave field $p(\mathbf{r}, t)$ obeys the wave equation:
\begin{equation}
	\left(\nabla^2 - \frac{1}{c^2}\frac{\partial^2}{\partial t^2}\right)
	p(\mathbf{r}, t)
	= -\frac{1}{c^2} Q(\mathbf{r}, t),
	\label{eq2}
\end{equation}
where $c$ is the wave speed. We assume the same temporal waveform of the source term $Q(\mathbf{r}, t)$ in both equations, while allowing different physical units.

To establish a formal connection between diffusion and wave phenomena, we introduce a virtual wave field $T_{\mathrm{virt}}(\mathbf{r}, t)$ defined as:
\begin{equation}
	\left(\nabla^2 - \frac{1}{c^2}\frac{\partial^2}{\partial t^2}\right)
	T_{\mathrm{virt}}(\mathbf{r}, t)
	= -\frac{1}{c^2} Q(\mathbf{r}, t).
	\label{eq3}
\end{equation}

The Fourier transform pair is defined as:
\begin{equation}
	\tilde{T}(\mathbf{r}, \omega)
	= \int_{-\infty}^{\infty} T(\mathbf{r}, t)e^{-i\omega t}\, dt,
	\label{eq4}
\end{equation}

\begin{equation}
	T(\mathbf{r}, t)
	= \frac{1}{2\pi} \int_{-\infty}^{\infty}
	\tilde{T}(\mathbf{r}, \omega)e^{i\omega t}\, d\omega,
	\label{eq5}
\end{equation}
where $\tilde{T}(\mathbf{r}, \omega)$ is the temperature field in frequency space.

Applying the Fourier transform to Eq.~\eqref{eq1} yields:
\begin{equation}
	\left(\nabla^2 - \sigma^2(\omega)\right)\tilde{T}(\mathbf{r}, \omega)
	= -\frac{1}{\kappa}\tilde{Q}(\mathbf{r}, \omega),
	\label{eq6}
\end{equation}
where $\sigma^2(\omega) \equiv i\omega/\alpha$. The complex wavenumber $\sigma(\omega)$ is chosen such that $\mathrm{Re}[\sigma(\omega)] > 0$, ensuring spatially decaying thermal diffusion waves.

Similarly, the frequency-domain form of Eq.~\eqref{eq3} reads:
\begin{equation}
	\left(\nabla^2 + k^2(\omega)\right)\tilde{T}_{\mathrm{virt}}(\mathbf{r}, \omega)
	= -\frac{1}{c^2}\tilde{Q}(\mathbf{r}, \omega),
	\label{eq7}
\end{equation}
with the real wavenumber $k(\omega) \equiv \omega/c$.
Eqs.~\eqref{eq6} and \eqref{eq7} are both Helmholtz-type equations; however, Eq.~\eqref{eq6} involves a complex wavenumber characteristic of diffusion, while Eq.~\eqref{eq7} describes propagating waves.

Motivated by the formal similarity between Eqs.~\eqref{eq6} and \eqref{eq7}, we introduce a correspondence between the thermal diffusion field and the virtual wave field via analytic continuation in the complex frequency plane. Specifically, replacing $\omega$ in Eq.~\eqref{eq7} by $-ic\sigma(\omega)$ leads to
\begin{equation}
	\tilde{T}(\mathbf{r}, \omega)
	= \frac{c^2}{\kappa}\,
	\tilde{T}_{\mathrm{virt}}(\mathbf{r}, -ic\sigma(\omega)),
	\label{eq8}
\end{equation}
which defines a mapping between the diffusion field and the virtual wave field in the frequency domain.

Equation~\eqref{eq8} should be regarded as a mathematically constructed correspondence rather than a physical equivalence. Transforming Eq.~\eqref{eq8} back to the time domain yields
\begin{equation}
	T(\mathbf{r}, t)
	= \frac{1}{2\pi} \int_{-\infty}^{\infty}
	\frac{c^2}{\kappa}
	\tilde{T}_{\mathrm{virt}}(\mathbf{r}, -ic\sigma(\omega))
	e^{i\omega t} \, d\omega,
	\label{eq9}
\end{equation}

where
\begin{equation}
	\tilde{T}_{\mathrm{virt}}(\mathbf{r}, -ic\sigma(\omega))
	= \int_{-\infty}^{\infty}
	T_{\mathrm{virt}}(\mathbf{r}, t')
	e^{-c\sigma(\omega)t'} \, dt'.
\end{equation}

Equation~\eqref{eq9} can be rewritten as a Fredholm integral equation of the first kind,
\begin{equation}
	T(\mathbf{r}, t)
	= \int_{-\infty}^{\infty}
	T_{\mathrm{virt}}(\mathbf{r}, t')\, K(t, t') \, dt',
	\label{eq10}
\end{equation}
where the kernel $K(t,t')$ enforces causality.

The kernel can be evaluated analytically as
\begin{equation}
	K(t, t')
	= \frac{c^2}{\kappa}
	\frac{t'}{2\sqrt{\pi}(\alpha t)^{3/2}}
	\exp\!\left(-\frac{c^2 t'^2}{4\alpha t}\right)
	H(t)\,H(t'),
	\label{eq11}
\end{equation}
where $H(\cdot)$ denotes the Heaviside step function.

The kernel explicitly enforces causality and reflects the irreversible nature of thermal diffusion, whereby the temperature field at time $t$ depends on contributions from all prior virtual-wave times $t' > 0$.
Eqs.~\eqref{eq10} and \eqref{eq11} connect the temperature signal to the virtual wave signal at the same spatial location $\mathbf{r}$ via a Fredholm integral equation of the first kind (Fig.~1a).

Due to the fact that the photothermal signal is discrete in time and space, Eq.~\eqref{eq10} can be discretized as
\begin{equation}
	\mathbf{T} = \mathbf{K}\mathbf{T}_{\mathrm{virt}},
	\label{eq12}
\end{equation}
where $\mathbf{T}$ and $\mathbf{T}_{\mathrm{virt}}$ denote the measured photothermal signal and the virtual wave signal at discrete time steps, respectively.

The inversion of Eq.~\eqref{eq12} is ill-conditioned because the matrix $\mathbf{K}$ is rank-deficient:
(1) thermal diffusion spreads heat in time, similar to how wave scattering smooths acoustic pressure signals, leading to nearly linearly dependent columns of $\mathbf{K}$;
(2) short-time features of the virtual wave are exponentially attenuated in the thermal response, analogous to high-frequency ultrasound attenuation in tissue;
(3) direct inversion amplifies measurement noise, producing unphysical oscillations in the reconstructed virtual wave.

The numerical methods for harmonic and multi-pulse excitations are presented in the following section.

\subsection{Numerical calculation for pulse excitation}

An iterative non-linear regularization technique, the alternating direction method of multipliers (ADMM), is used to reconstruct $\mathbf{T}_{\mathrm{virt}}$.
ADMM enables the incorporation of sparsity as prior information, reducing artificial oscillations in the solution.

We assume sparsity in the virtual wave field due to a limited number of material interfaces. The inverse problem can be formulated as
\begin{equation}
	\min_{\mathbf{T}_{\mathrm{virt}}}
	\frac{1}{2}\|\mathbf{K}\mathbf{T}_{\mathrm{virt}} - \mathbf{T}\|_2^2
	+ \lambda_{\mathrm{ADMM}} \|\mathbf{T}_{\mathrm{virt}}\|_1,
	\label{eq13}
\end{equation}
where the $\ell_1$-norm promotes sparsity in $\mathbf{T}_{\mathrm{virt}}$. The regularization parameter $\lambda_{\mathrm{ADMM}}$ is determined using the L-curve criterion. The discretized version for two spatial dimensions $x$ and $y$ is given in Eq.~\eqref{eq12}, where
$\mathbf{T} \in \mathbb{R}^{N_t \times N_x \times N_y}$,
$\mathbf{K} \in \mathbb{R}^{N_t \times N_{t'}}$, and
$\mathbf{T}_{\mathrm{virt}} \in \mathbb{R}^{N_{t'} \times N_x \times N_y}$.

The entries of the discrete kernel $\mathbf{K} = [K_{i,j}] \in \mathbb{R}^{N_t \times N_{t'}}$ are computed as
\begin{equation}
	K(i,j)
	= \frac{\tilde{c}^{\,2}}{\kappa}
	\frac{(j-1)}{2\sqrt{\pi}(\tilde{\alpha} i)^{3/2}}
	\exp\!\left(
	-\frac{\tilde{c}^{\,2}(j-1)^2}{4\tilde{\alpha} i}
	\right),
	\label{eq14}
\end{equation}
where
$\tilde{c} = c \Delta t / \Delta z$
and
$\tilde{\alpha} = \alpha \Delta t' / \Delta z^2$
are dimensionless parameters representing normalized wave speed and thermal diffusivity, respectively. Here $\Delta t$, $\Delta t'$, and $\Delta z$ denote the time, virtual time, and spatial step sizes.
\subsection{Numerical calculation for harmonic wave excitation}

We perform the singular value decomposition (SVD) of the kernel matrix:
\begin{equation}
	\mathbf{K} = \mathbf{U}\mathbf{\Sigma}\mathbf{V}^T,
	\label{eq15}
\end{equation}
where $\mathbf{U} = (u_1,\dots,u_N)$ and $\mathbf{V} = (v_1,\dots,v_N)$ are orthonormal matrices, and
$\mathbf{\Sigma} = \mathrm{diag}(\sigma_1,\dots,\sigma_N)$ with $\sigma_1 \ge \sigma_2 \ge \cdots \ge \sigma_N \ge 0$. The formal solution of the inverse problem is given by
\begin{equation}
	\mathbf{T}_{\mathrm{virt}}
	= \sum_{n=1}^{N}
	\frac{u_n^T \mathbf{T}}{\sigma_n}\, v_n.
	\label{eq16}
\end{equation}

Because thermal diffusion suppresses high temporal frequencies exponentially, the singular values decay approximately as
$\sigma_n \sim \exp(-\beta n)$, where $\beta \propto \sqrt{\alpha}/\Delta t$.
As a result, small singular values dominate the inversion and amplify noise.

To stabilize the inversion, we introduce a truncation index $r$ and define the truncated inverse operator as
\begin{equation}
	\mathbf{K}_r^{-1}
	= \sum_{n=1}^{r}
	\frac{1}{\sigma_n} v_n u_n^T,
	\label{eq17}
\end{equation}
and the reconstructed virtual wave field becomes
\begin{equation}
	\mathbf{T}_{\mathrm{virt}}^{(r)}
	= \sum_{n=1}^{r}
	\frac{u_n^T \mathbf{T}}{\sigma_n}\, v_n.
	\label{eq18}
\end{equation}
which removes components associated with $\sigma_n \ll \sigma_r$.
The diffusion kernel in Eq.~\eqref{eq11} acts as a temporal Gaussian filter with characteristic width
\begin{equation}
	\delta t \sim \frac{4\alpha t}{c^2}.
\end{equation}

Thus, the effective temporal bandwidth of recoverable virtual-wave components satisfies
\begin{equation}
	\omega_{\max} \sim \frac{c^2}{4\alpha t}.
\end{equation}

The truncation index $r$ therefore corresponds to a diffusion-limited temporal resolution, analogous to the diffraction limit in wave imaging. Equivalently, the Picard condition can be written as
\begin{equation}
	|u_n^T \mathbf{T}| \ll \sigma_n \quad \text{for large } n.
\end{equation}

Stable inversion requires truncation before the noise floor intersects the singular spectrum. From Eq.~\eqref{eq11}, short-time virtual-wave components are attenuated as
\begin{equation}
	\exp\!\left(-\frac{c^2 t'^2}{4\alpha t}\right).
\end{equation}

The smallest resolvable virtual time satisfies
\begin{equation}
	t'_{\min} \sim \sqrt{\frac{4\alpha t}{c^2}\ln\!\left(\frac{1}{\mathrm{SNR}}\right)}.
\end{equation}

Hence, truncated SVD (T-SVD) does not merely regularize numerically; it enforces the thermodynamic irreversibility constraint inherent in diffusion. The T-SVD inversion operator reads
\begin{equation}
	\mathbf{T}_{\mathrm{virt}}^{(r)}
	= \mathbf{V}_r \mathbf{\Sigma}_r^{-1} \mathbf{U}_r^T \mathbf{T},
	\label{eq19}
\end{equation}
where the subscript $r$ denotes restriction to the dominant singular subspace.
\subsection{Relationship between excitation waveform and inversion strategy}

From a signal representation perspective, the suitability of ADMM or T-SVD is closely related to the temporal structure of the thermal excitation. A pulse-type excitation generates a thermal response that is localized in virtual time, i.e.,
\begin{equation}
	T_{\mathrm{virt}}(t')
	= \sum_{k=1}^{K} a_k \delta(t' - t_k'),
	\label{eq20}
\end{equation}
or, more generally, a piecewise-smooth signal with a small number of dominant events associated with defect activation. When discretized, such signals admit a sparse representation in the time domain, meaning that most entries of $T_{\mathrm{virt}}$ are close to zero. For this class of signals, the reconstruction naturally leads to an $\ell_1$-regularized formulation, since the solution satisfies
\begin{equation}
	\|T_{\mathrm{virt}}\|_0 \ll N_{t'},
	\label{eq21}
\end{equation}
where $\|\cdot\|_0$ denotes the number of non-zero components. ADMM explicitly promotes this sparsity by minimizing the convex surrogate $\|T_{\mathrm{virt}}\|_1$, allowing sharp temporal transitions to be preserved even under strong diffusion-induced blurring. Consequently, burst-induced non-harmonic responses are reconstructed more faithfully using ADMM, with reduced ringing artifacts and improved localization of defect-related events. In contrast, under harmonic modulated excitation, the thermal response can be approximated as
\begin{equation}
	T(t) = \mathrm{Re}\{A(\omega)e^{i\omega t}\},
	\label{eq22}
\end{equation}
which, after inversion, yields a virtual-wave field dominated by a small number of narrowband frequency components. In the SVD framework, such signals exhibit a strong projection onto the leading singular vectors $u_n$ ($n=1,\dots,r$), i.e.,
\begin{equation}
	|u_n^T \mathbf{T}| \gg 0 \quad \text{for } n \le r, 
	\qquad
	|u_n^T \mathbf{T}| \approx 0 \quad \text{for } n > r.
	\label{eq23}
\end{equation}
thereby satisfying the Picard condition. The energy of harmonic excitation is thus concentrated in the dominant singular subspace, making T-SVD an efficient and physically consistent inversion method. The truncation index r effectively removes diffusion-dominated high-order modes, while retaining the coherent harmonic response with minimal bias. Therefore, ADMM and T-SVD are not interchangeable numerical tools, but rather correspond to different assumptions on the temporal sparsity or spectral compactness of the virtual thermal wave field. Pulse-type, non-harmonic excitations favor sparse, time-localized reconstructions and are best handled by ADMM, whereas harmonic excitations generate spectrally compact responses that align naturally with the low-rank structure exploited by T-SVD.

\section{Numerical simulation results}
The first example considers the most commonly used Dirac pulse excitation. In practice, pulsed excitation always has a finite temporal width. Therefore, to better approximate real experiments, the heat excitation is modeled as a square pulse with a duration of 2 ms. The tested material is stainless steel ($40~\mathrm{mm} \times 4~\mathrm{mm}$, where the depth and size of defects are $1~\mathrm{mm}$ and $10~\mathrm{mm}$), with thermal conductivity $\kappa = 16.2~\mathrm{W\,m^{-1}K^{-1}}$, density $\rho = 7900~\mathrm{kg\,m^{-3}}$, and heat capacity $c_p = 477~\mathrm{J\,kg^{-1}K^{-1}}$.

The temperature variations between defect and sound (non-defect) regions are shown in Fig.~1b. According to the artificially defined tomographic relationship, time is often related to depth via the diffusion length
\begin{equation}
	L_{\mathrm{diff}} = \sqrt{\alpha t}.
\end{equation}

\begin{figure*}
	\centering
	\includegraphics[width=1.0\textwidth]{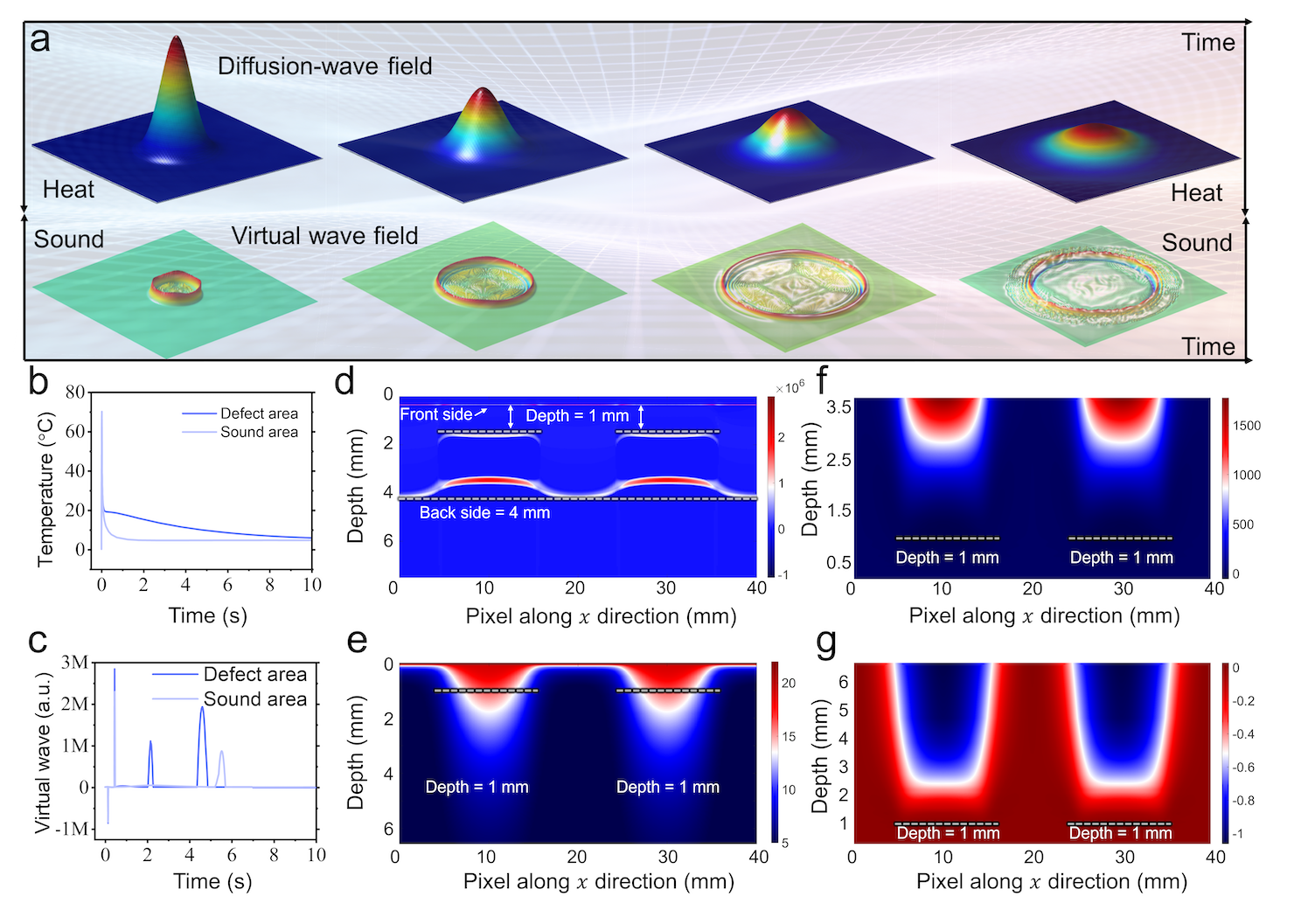}
	\caption{The transformation from diffusion-wave field to virtual-wave field for Dirac pulse excitation. a Schematic image of virtual wave method. b Temperature variation between defect area and sound area. c Virtual-wave variation between defect area and sound area. d Virtual-wave tomography results. e Time-domain photothermal tomography results. f Amplitude-based tomography results. g Phase-based tomography results. }\label{fig1}
\end{figure*}

However, the defect appears in the first frame and gradually diffuses with time/depth (Fig.~\ref{fig1}e), which is attributed to the Markovian nature of heat diffusion. This demonstrates that time-domain signals cannot be directly interpreted as depth-resolved tomographic information, since heat propagation is not bounded by a finite propagation velocity.

Following Zhu et al.~\cite{ref14}, the frequency in the one-dimensional analytical solution under sinusoidal excitation can be treated as a continuous variable. Thus, the frequency-domain response of pulsed thermography can be interpreted as a superposition of harmonic components with different frequencies, where the thermal diffusion length is given by
\begin{equation}
	\mu = \sqrt{\frac{\alpha}{\pi f}}.
\end{equation}

The amplitude tomograms exhibit depth-dependent variations (Fig.~\ref{fig1}f). However, since heat transport is governed by gradient-driven diffusion, the amplitude tomograms do not exhibit sharp boundaries. Moreover, the reconstructed depths deviate from the true physical depth when using an automatically calibrated colormap.

Similar behavior is observed in phase tomograms (Fig.~\ref{fig1}g). Although the main defect distribution in the phase image is closer to the true depth compared to the amplitude image, photothermal tomography remains limited by blurred boundaries and an artificially defined depth–frequency relationship~\cite{ref32,ref33,ref34}.
Virtual-wave transformation in this work thoroughly changes this situation. Based on the derived relation (Eqs.~\eqref{eq10} and~\eqref{eq11}) between thermal diffusion-wave equation and wave equation, the photothermal signal can be transformed into the wave-like signal using ADMM algorithm. The transformed results are shown in Fig.~\ref{fig1}c. It is possible to find that the defect area has a clearly reflected echo. Furthermore, the sound area presents a reflected echo caused by the back side. The thickness of the sample is impossible to achieve in any photothermal signal because the direction in diffusion-wave field is one-way~\cite{ref35}. The clear and gradient-unrelated tomogram is obtained by reconstructing the entire virtual-wave field (Fig.~\ref{fig1}d). Different from the previous work in photothermal reconstruction~\cite{ref36,ref37}, we construct a real wavefront in photothermal signals.

\begin{figure*}
	\centering
	\includegraphics[width=1.0\textwidth]{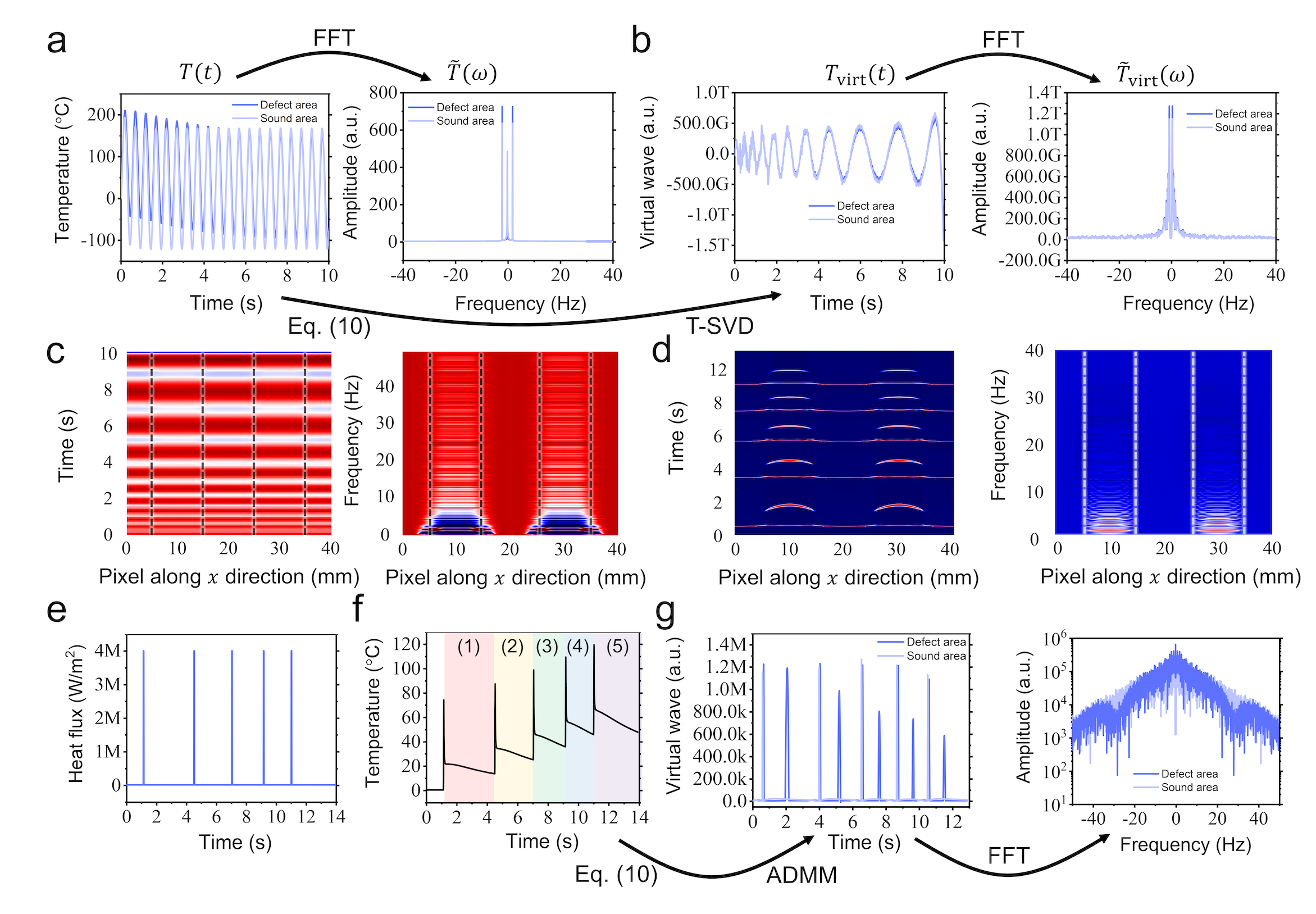}
	\caption{The transformation from diffusion-wave field to virtual-wave field for modulated excitation. a Temperature signals under lock-in excitations in the time and frequency domain. b Reconstructed virtual wave signals under lock-in excitations in the time and frequency domain. c Reconstructed virtual wave field under lock-in excitations in the time and frequency domain. d Reconstructed virtual wave field under chirp-pulsed excitations in the time and frequency domain. e The excitation signal. f The temperature response signal under chirp-pulsed excitations. g Reconstructed virtual wave signals under chirp-pulsed excitations in the time and frequency domain.}\label{fig2}
\end{figure*}

Then, we try to use virtual wave methods to reconstruct the photothermal signal under modulated excitations, including harmonic wave~\cite{ref38} and chirp-pulsed signals~\cite{ref39}. For the harmonic wave, the modulation frequency is set to 1 Hz. The temperature variation between defect area and sound area is shown in Fig.~\ref{fig2}a. The first-order linear fitting was used to remove the DC component. According to the amplitude signals (Fig.~\ref{fig2}a), it is clear to find that the peak is located at 1 Hz. Different from pulsed excitation, truncated singular value decomposition (T-SVD) was employed to numerically calculate the inverse problem for harmonic wave excitations. The reconstructed virtual wave signals are shown in Fig.~\ref{fig2}b. Then, the virtual-wave signals were transformed into the frequency domain. It is also obvious that the peak located around 0.5 Hz since the T-SVD cannot provide a perfect inversion (Fig.~\ref{fig2}b). The entire virtual wave field map in the time domain is shown in Fig.~\ref{fig2}c. By transforming the virtual wave field into the frequency domain, it is clear to find two subsurface defects (Fig.~\ref{fig2}c). These results demonstrate the accuracy of the derived virtual wave equations.
Compared to harmonic signals, the clear advantage of shirt pulse capable of maintaining a flat power spectrum over a very broad bandwidth, even under diffusive attenuation losses. The chirped-pulse thermal-diffusion-wave radar technique has the advantages of a highly compressed output with negligible side-lobe power distribution led to a much improved signal-to-noise ratio (SNR) and depth profiling capability. In this work, a fixed-width excitation pulse in the frequency range between 0.2 and 0.6 Hz, a chirp duration of 12 s, and pulse duration of the excitation chirp of 2 ms was used (Fig.~\ref{fig2}e). The ADMM was employed to numerically calculate virtual wave signals. Of note, since heat diffusion is a linear superposition system, we separated the chirp-pulsed excitation based on the beginning position of each peak before using ADMM algorithm (Fig.~\ref{fig2}f). The reconstructed virtual wave signals in the time and frequency domain are shown in Fig.~\ref{fig2}g. It is clear to find the decreasing magnitude echoes in the defect area. Compared with the single-frequency harmonic wave excitation, chirp-pulsed excitation exhibits a broad bandwidth (Fig.~\ref{fig2}g). The reconstructed virtual wave field in the time and frequency domain is shown in Fig.~\ref{fig2}d. It is possible to find that the chirp-pulsed signals exhibit highly compressed output with negligible side-lobe power distribution.

\section{Photothermal experiment results}
\subsection{Experimental setup}
The resulting temperature rise was recorded with a spatial resolution of
$\Delta x, \Delta y = 52~\mu\mathrm{m}$
at a frame rate of
$f_{\mathrm{cam}} = 100~\mathrm{Hz}$
using a cooled mid-wave infrared (MWIR) camera (ImageIR 9300, Infratec). To achieve coaxial alignment of the laser illumination and IR detection, a dichroic mirror was used, which reflects the $\lambda = 940~\mathrm{nm}$ laser while transmitting in the MWIR spectral range. The region of interest (ROI) was sampled over $n_m = 403$ measurement points arranged on a grid with spacing
$r_d = 0.743~\mathrm{mm}$.
\begin{figure*}
	\centering
	\includegraphics[
	width=1.0\textwidth,
	height=0.88\textheight,
	keepaspectratio
	]{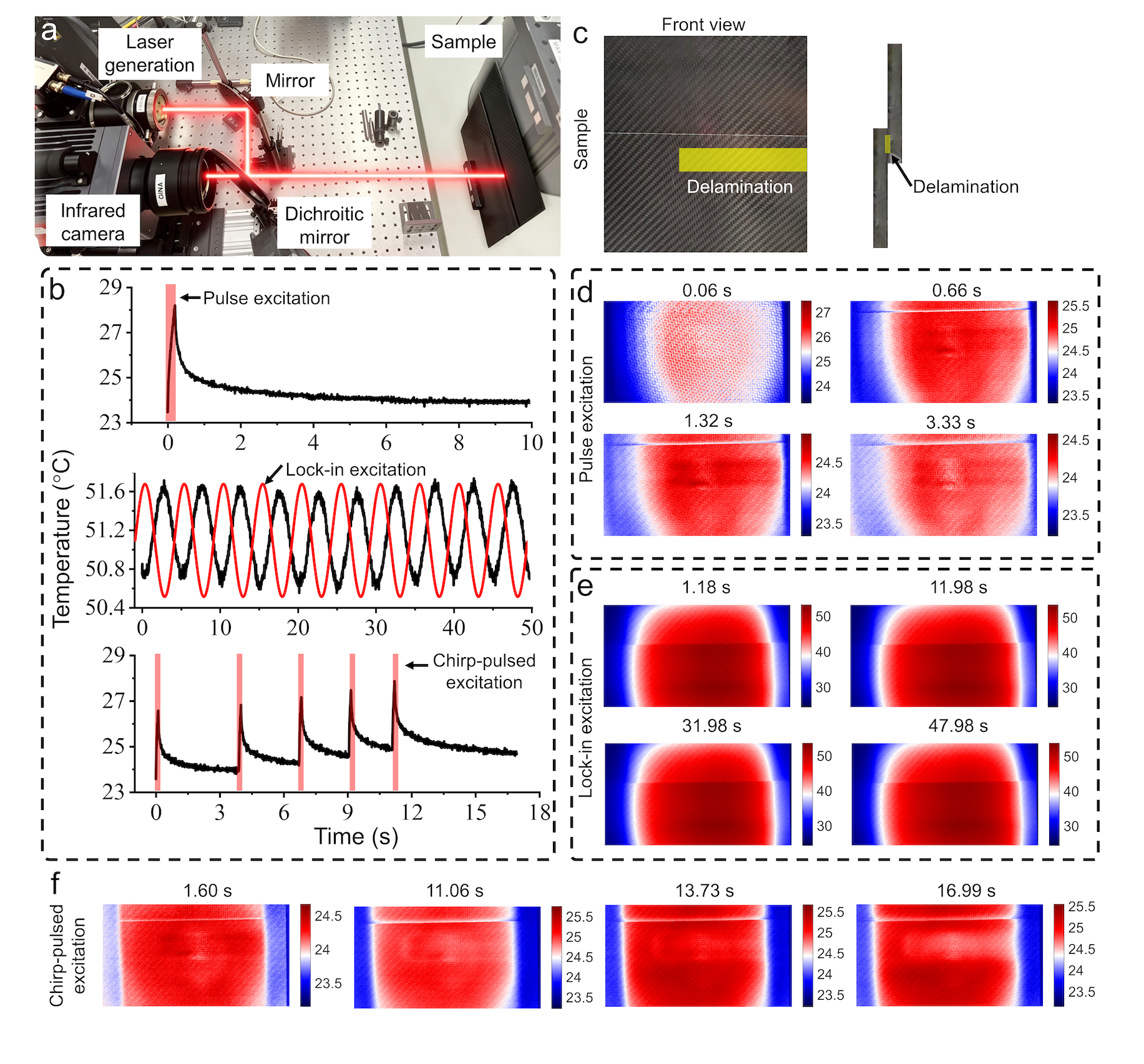}
	\caption{The photothermal experimental setup and results. a Experimental setup. b The excitation signal waveform. c The photograph of tested sample. d Original results from pulse excitation. e Original results from lock-in excitation. f Original results from chirp-pulsed excitation.}\label{fig3}
\end{figure*}
\subsection{Experimental results}
\begin{figure*}
	\centering
	\includegraphics[
	width=1.0\textwidth,
	height=0.88\textheight,
	keepaspectratio
	]{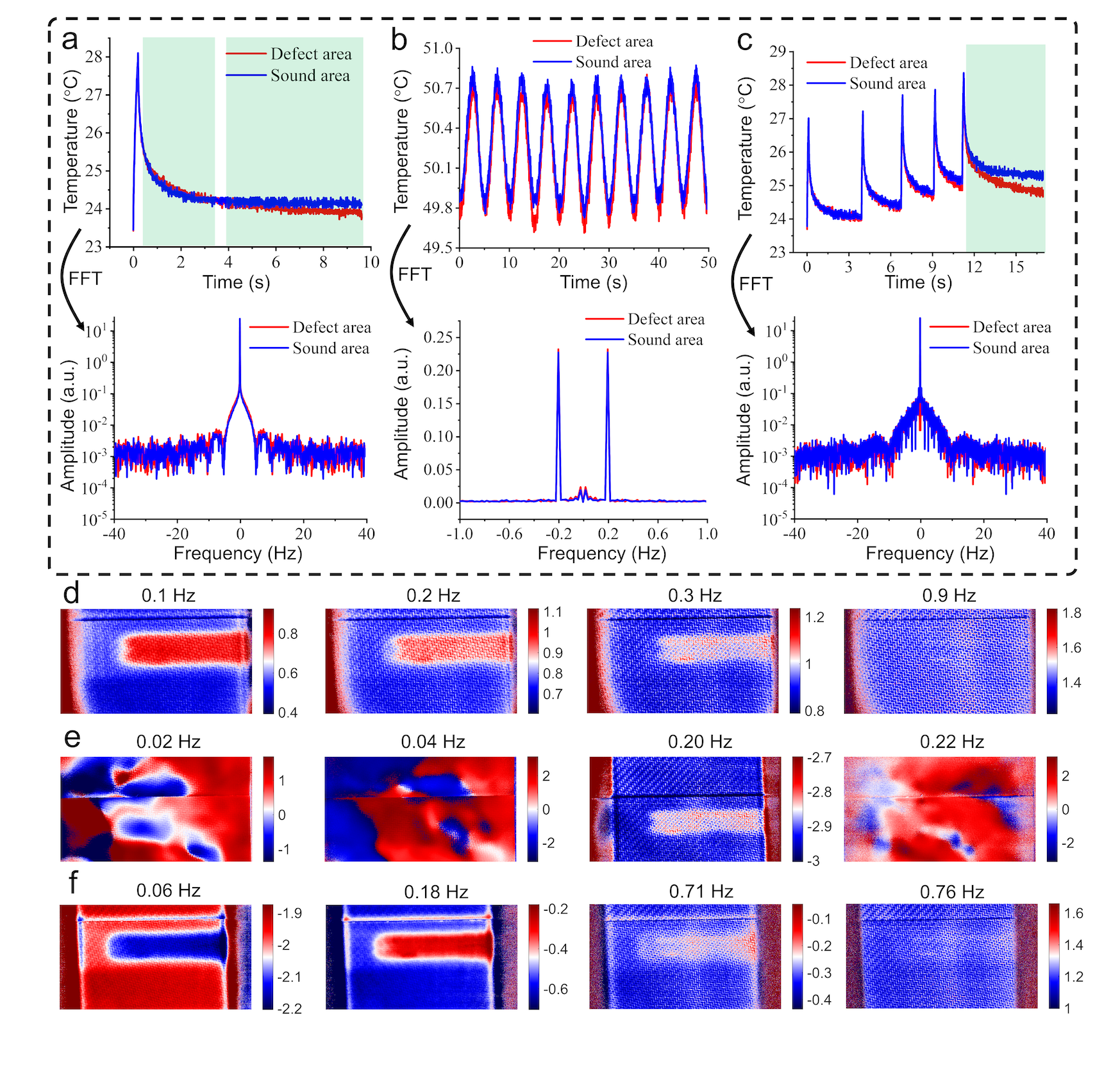}
	\caption{The signal analysis for photothermal experiments. a Temporal and frequency-domain signal under pulse excitation. b Temporal and frequency-domain signal under lock-in excitation. c Temporal and frequency-domain signal under chirp-pulsed excitation. d Frequency-domain results from pulse excitation. e Frequency-domain results from lock-in excitation. f Frequency-domain results from chirp-pulsed excitation.}\label{fig4}
\end{figure*}
To validate the proposed generalized virtual wave theory, photothermal experiments were conducted. The laser beam generated by a laser source was reflected by a mirror and projected onto the sample surface, forming a spatially modulated rectangular heating region. The temporal profile of the excitation was controlled by a signal generator, enabling arbitrary waveform modulation. The resulting surface temperature evolution and spatial distribution were recorded using an infrared camera. A dichroic mirror was used to establish a $90^\circ$ optical path while preventing the reflected laser beam from directly entering the infrared camera (Fig.~\ref{fig3}a). The excitation waveforms can be classified into three types: pulse excitation~\cite{ref40}, lock-in excitation~\cite{ref41}, and chirp-pulsed excitation~\cite{ref42} (Fig.~3b). The pulse duration of the pulse excitation is 120 ms. The modulation frequency of the lock-in excitation is 0.2 Hz. For the chirp-pulsed excitation, a frequency-swept excitation in the range of 0.2–0.6 Hz is used, with a chirp duration of 16 s and a pulse width of 100 ms. The tested sample consists of two carbon fiber-reinforced polymer (CFRP) laminates with a Teflon insert sandwiched between them (Fig.~\ref{fig3}c). The thermal properties of the CFRP laminate are: thermal conductivity $k = 0.8~\mathrm{W\,m^{-1}K^{-1}}$, heat capacity $c_p = 1200~\mathrm{J\,kg^{-1}K^{-1}}$, and density $\rho = 1600~\mathrm{kg\,m^{-3}}$. For Teflon, the thermal conductivity, heat capacity, and density are $0.25~\mathrm{W\,m^{-1}K^{-1}}$, $1050~\mathrm{J\,kg^{-1}K^{-1}}$, and $2170~\mathrm{kg\,m^{-3}}$, respectively.

The original results under pulse excitation are shown in Fig.~\ref{fig3}d. It is possible to observe the obscured delamination at 0.66 s and 1.32 s. The defect contrast becomes lower at 3.33 s. However, it is still not clear to find the exact boundary or shape of this delamination defect. The original results under lock-in excitation are shown in Fig.~\ref{fig3}e. Unfortunately, it is difficult to detect the delamination defect in the time domain. The original results under chirp-pulsed excitation are shown in Fig.~\ref{fig3}f. It is obvious to observe the delamination with time increasing. In addition, the defect contrast under chirp-pulsed excitation is higher than that under pulse excitation.

To further analyze the defect feature, the temporal and frequency-domain signals were extracted, as shown in Fig.~\ref{fig4}. According to the time-domain signals of defect and sound area (Fig.~\ref{fig4}a), the temperature exhibits difference in two regions before 2.7 s and after 4 s. The frequency-domain signals show that the signal energy significantly decreases after 5 Hz. As for the lock-in excitation, it is no obvious difference between defect and sound area in the time domain (Fig.~\ref{fig4}b). Therefore, the defect cannot be detected in the previous temporal results (Fig.~\ref{fig3}e). The frequency-domain signal demonstrates a sharp peak at 0.2 Hz, which is also the modulated frequency of the lock-in heat source. The chirp-pulsed signals between defect and sound area present obvious difference after the last pulse excitation (Fig.~\ref{fig4}c). It is noted that the frequency-domain signals under chirp-pulsed excitation demonstrate wider spectrum than that under pulse excitation. This is because the chirp excitation sweeps through a range of frequencies during the excitation period, it injects energy over a broader frequency band than a single short pulse, resulting in a wider spectrum in the frequency domain.

The frequency-domain results under pulse excitation are shown in Fig.~\ref{fig4}d. The delamination area in the frequency domain becomes more obvious than that in the time domain (Fig.~\ref{fig3}d). The rectangular shape of the inserted Teflon can be detected at 0.1, 0.2, and 0.3 Hz. For results under lock-in excitation, only 0.2 Hz frequency component provides effective information for the tested sample and delamination defect (Fig.~\ref{fig4}e). As for results under chirp-pulsed excitation, the delamination can be detected before 0.71 Hz (Fig.~\ref{fig4}f). However, it disappears at 0.76 Hz. Combining the knowledge of frequency-domain signal feature analysis (Fig.~\ref{fig4}c), chirp-pulsed excitation has sufficient energy within 10 Hz. Therefore, the thermogram at 0.76 Hz provides accurate information that the delamination interface is related to this frequency component.
\begin{figure*}
	\centering
	\includegraphics[width=1.0\textwidth]{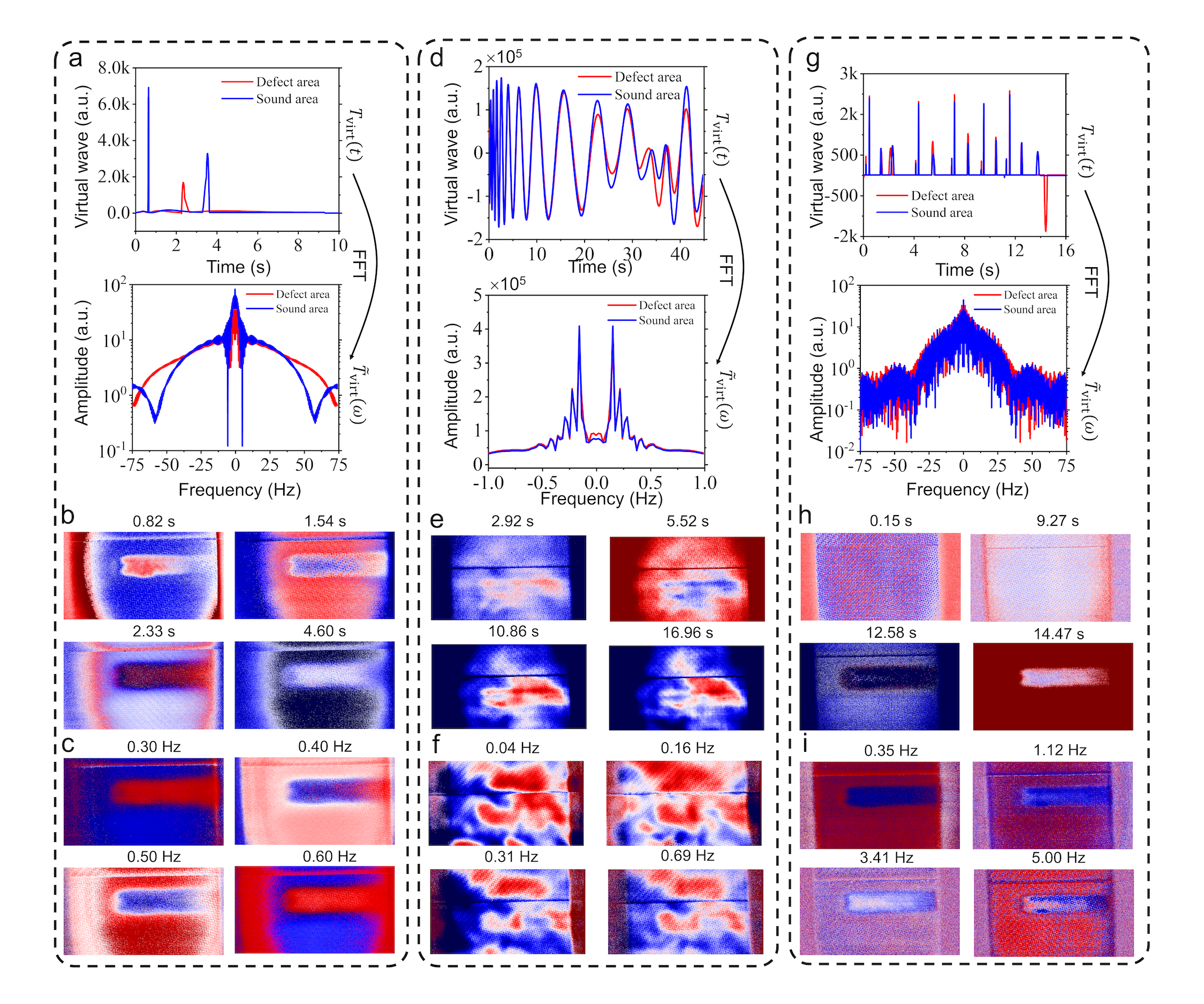}
	\caption{The virtual wave reconstruction of photothermal experiments. a Temporal and frequency-domain signal under pulse excitation. b Time-domain virtual wave images under pulse excitation. c Frequency-domain virtual wave images under pulse excitation. d Temporal and frequency-domain signal under lock-in excitation. e Time-domain virtual wave images under lock-in excitation. f Frequency-domain virtual wave images under lock-in excitation. g Temporal and frequency-domain signal under chirp-pulsed excitation. h Time-domain virtual wave images under chirp-pulsed excitation. i Frequency-domain virtual wave images under chirp-pulsed excitation.}\label{fig5}
\end{figure*}
Figure~\ref{fig5} presents the virtual‑wave reconstruction results obtained from photothermal experiments under pulse, lock‑in, and chirp‑pulsed excitations, providing a direct experimental validation of the generalized virtual‑wave framework. In all three cases, the reconstructed virtual‑wave signals and images exhibit wave‑like characteristics that are absent in the original thermal diffusion data, demonstrating the successful transformation from a diffusive field to a propagative representation. For pulse excitation (Fig.~\ref{fig5}a-c), the reconstructed virtual‑wave signals show distinct, temporally localized peaks in the defect region that are clearly separated from those in the sound region. These sharp echoes are characteristic of wave reflections from material interfaces and stand in strong contrast to the broadened, monotonically decaying temperature responses typically observed in time‑domain thermography. The corresponding time‑domain virtual‑wave images (Fig.~\ref{fig5}b) reveal a well‑defined subsurface feature with clear spatial boundaries appearing at specific virtual times, while the frequency‑domain virtual‑wave images (Fig.~\ref{fig5}c) exhibit consistent defect contrast over a limited frequency range. This behavior indicates that the pulse excitation generates a sparse virtual‑wave response dominated by a small number of reflection events, which is well suited to ADMM‑based inversion and enables accurate localization of the delamination interface. Under lock‑in excitation (Fig.~\ref{fig5}d-f), the temporal virtual‑wave signals exhibit a quasi‑periodic structure inherited from the harmonic heating, with a pronounced difference between defect and sound areas becoming evident only after virtual‑wave reconstruction. While the original thermal signals in the time domain show little contrast, the virtual‑wave representation converts phase‑shifted diffusive responses into wave‑like oscillations whose amplitudes and phases vary spatially. The time‑domain virtual‑wave images (Fig.~\ref{fig5}e) display defect‑related patterns that are difficult to identify in conventional lock‑in thermography, and the frequency‑domain virtual‑wave images (Fig.~\ref{fig5}f) highlight the delamination most clearly at specific frequencies corresponding to the effective penetration depth of the harmonic excitation. These results confirm that, although lock‑in excitation is intrinsically narrowband, the virtual‑wave transformation enables wave‑based interpretation and depth‑selective imaging through frequency discrimination. The chirp‑pulsed excitation results (Fig.~\ref{fig5}g-i) further illustrate the advantages of combining broadband thermal excitation with the generalized virtual‑wave framework. The reconstructed virtual‑wave signals show a sequence of decaying echoes with a substantially broader frequency content than those obtained under pulse or lock‑in excitation. In the time‑domain virtual‑wave images (Fig.~\ref{fig5}h), the defect region emerges with high contrast at well‑defined virtual times, while the frequency‑domain images (Fig.~\ref{fig5}i) reveal consistent defect visibility over a wide frequency range. This broadband response not only improves signal‑to‑noise ratio through spectral averaging, but also provides richer depth‑related information, as different frequency components correspond to different effective propagation lengths in the virtual‑wave domain. As a result, chirp‑pulsed excitation offers superior robustness and depth interpretability compared with single‑frequency lock‑in or single‑pulse heating.

For comparing the image quality (defect contrast) between different results and methods, the quantitative metrics, the contrast-to-noise ratio (CNR), was employed, which is a commonly used index in NDE fields. The mathematical expression of CNR can be written as~\cite{ref43,ref44}:
The contrast-to-noise ratio (CNR) is defined as
\begin{equation}
	\mathrm{CNR} = \left| \frac{\mu_d - \mu_s}{\sigma_s} \right|,
	\label{eq24}
\end{equation}
where $\mu_d$ and $\mu_s$ denote the mean values of the defect and sound regions, respectively, and $\sigma_s$ is the standard deviation of the sound region.

Table~I provides a quantitative comparison of defect detectability for different excitation waveforms by reporting the maximum CNR in both the time and frequency domains, before and after virtual wave reconstruction. For all three excitation schemes, the virtual wave transformation leads to a significant improvement in CNR compared with the raw thermographic data, confirming the effectiveness of the proposed generalized virtual wave framework in mitigating diffusion-induced blurring and background noise. The improvement is particularly pronounced in the time domain for pulse and chirp-pulsed excitations, where the CNR increases by several orders of magnitude after virtual wave reconstruction. This indicates that diffusive thermal responses are successfully transformed into wave-like signals with temporally localized reflection events, enabling a substantially improved separation between defect and sound regions. In contrast, lock-in excitation exhibits relatively low CNR values in both raw and virtual-wave-reconstructed results. This behavior can be attributed to its inherently narrowband nature and limited depth sensitivity; nevertheless, a consistent increase in CNR is still observed after virtual wave processing, demonstrating the robustness of the proposed method even under harmonic excitation. In the frequency domain, the CNR enhancement is more moderate, reflecting the fact that frequency-domain representations are dominated by specific effective frequency components rather than temporally resolved reflection events.
\begin{table*}[htbp]
	\caption{Quantitative comparison of CNR performance for different excitation waveforms in time and frequency domains before and after virtual wave reconstruction.}
	\label{tab1}
	\centering
	
	\begin{tabular}{c c c c @{\hspace{0.1em}} c c c}
		
		\hline
		
		& \multicolumn{3}{c}{Raw}
		& \multicolumn{3}{c}{Virtual wave} \\
		
		\cline{2-4}\cline{5-7}
		
		Waveform
		& Pulse
		& Lock-in
		& Chirp-pulsed
		& Pulse
		& Lock-in
		& Chirp-pulsed \\
		
		\hline
		
		Time-domain
		& 1.17
		& 0.03
		& 4.20
		& 34.87
		& 2.50
		& 10651.57 \\
		
		Frequency-domain
		& 4.30
		& 3.29
		& 16.85
		& 12.88
		& 2.70
		& 2.96 \\
		
		\hline
		
	\end{tabular}
\end{table*}

\subsection{Comparison between existing photothermal coherence tomography techniques}
\begin{figure*}
	\centering
	\includegraphics[width=1.0\textwidth]{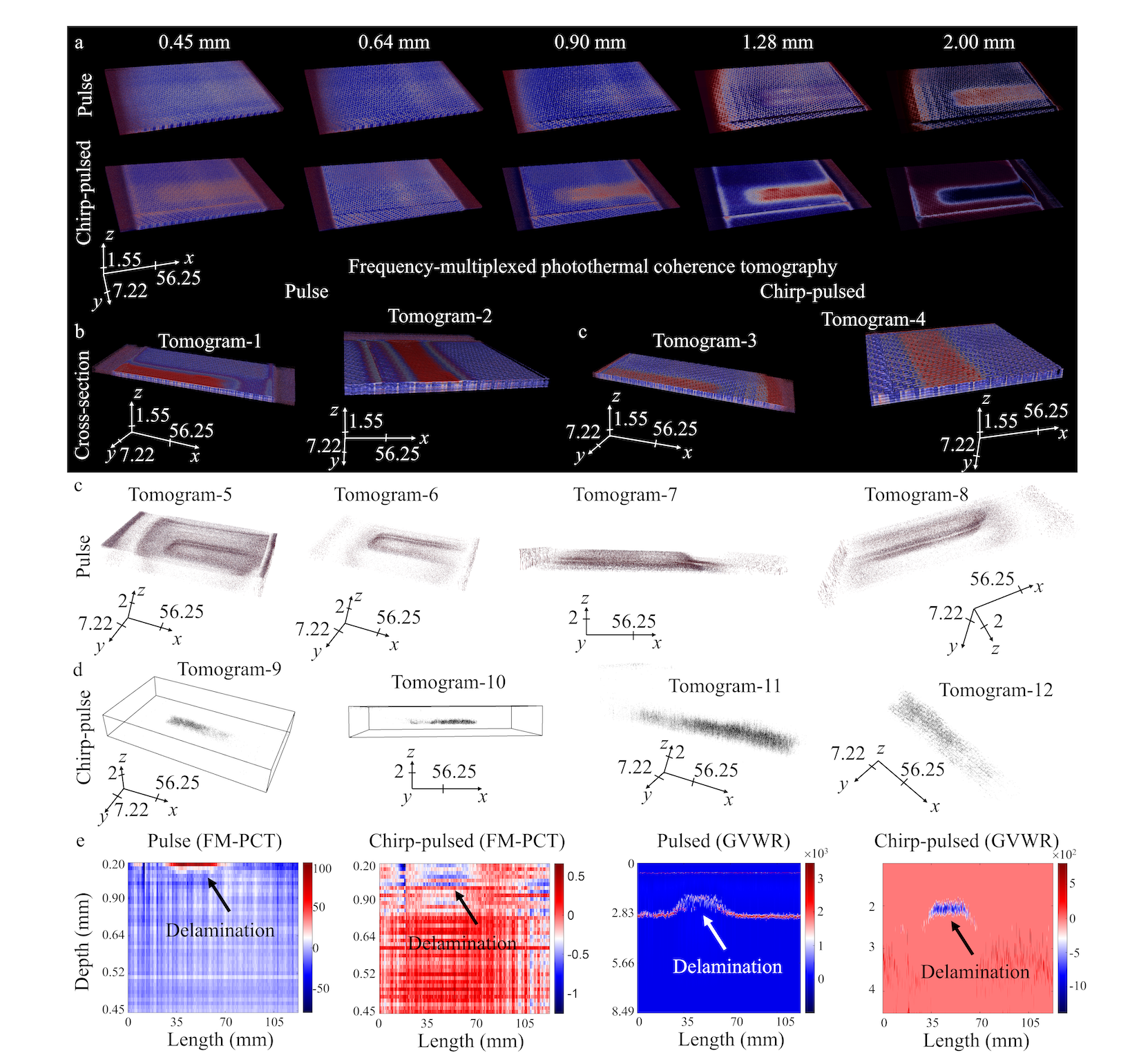}
	\caption{The comparison between generalized virtual wave reconstruction (GVWR) and frequency multiplexed photothermal coherence tomography (FM-PCT) technique. a Photothermal tomography results from FM-PCT along various depth layers for pulse and chirp-pulsed excitations. b Cross-sectional results from FM-PCT for pulse and chirp-pulsed excitations. c Photothermal tomography results from GVWR for pulse excitation. d Photothermal tomography results from GVWR for chirp-pulsed excitation. e Depth Slices for FM-PCT and GVWR under pulse and chirp-pulsed excitations.}\label{fig6}
\end{figure*}

Frequency-multiplexed photothermal coherence tomography (FM-PCT)~\cite{ref14} is an emerging technique in infrared thermography for retrieving 3D subsurface structures of tested samples. Compared with the enhanced truncated-correlation photothermal coherence tomography (eTC-PCT) technique~\cite{ref13}, FM-PCT does not require a modulated chirp-pulsed excitation or a high-frame-rate infrared camera. It shows excellent performance for pulse excitation or reconstructed pulse-like signals, such as linear scanning thermography. The photothermal tomography results processed by FM-PCT and generalized virtual wave reconstruction (GVWR) are shown in Fig.~\ref{fig6}. The tomograms at different depth layers are presented in Fig.~6a. The delamination appears at approximately 2 mm under pulse excitation. However, the chirp-pulsed results exhibit spectral aliasing, although they provide a higher signal-to-noise ratio (or defect contrast) compared with pulse excitation. The cross-sectional tomograms are shown in Fig.~\ref{fig6}b. According to slices 1–4, the delamination mainly exists in a thin region near the bottom of the sample. The GVWR results for pulse excitation are shown in Fig.~\ref{fig6}c, where the delamination is clearly visible and exhibits a continuous profile. The GVWR results for chirp-pulsed excitation are shown in Fig.~\ref{fig6}d, where the delamination exhibits extremely high contrast, appearing as a well-defined structure. To further quantitatively compare FM-PCT and GVWR, depth slices along the delamination and its surrounding region were extracted, as shown in Fig.~\ref{fig6}e. Consistent with previous analysis, the delamination exists only in a thin depth region. The chirp-pulsed results exhibit spectral aliasing, as the delamination appears at multiple depth locations and becomes discontinuous. In contrast, GVWR results for both pulse and chirp-pulsed excitation exhibit high accuracy and strong defect contrast. It is further noted that the depth mapping in infrared thermography is inherently nonlinear due to diffusive transport, regardless of whether time-domain or frequency-domain representations are used. For example, the empirical time-domain depth relation $z = 2\sqrt{\alpha t}/{\sqrt{\pi}}$
was reported by Mandelis et al.~\cite{ref3}, while the frequency-domain thermal diffusion length $\mu = \sqrt{{\alpha}/{\pi f}}$ was provided by Zhu et al.~\cite{ref14}. This explains why the delamination appears only within a thin effective depth region. However, in GVWR, the depth relationship follows a linear mapping $z = {ct}/{2}$, analogous to ultrasonic wave propagation, where $c$ is the virtual wave speed. Therefore, GVWR can be regarded as a true tomographic reconstruction framework compared with conventional photothermal coherence tomography techniques.

\section{Conclusion}
In this work, we have developed a generalized virtual-wave photothermal tomography framework that establishes a rigorous and unified transformation from thermal diffusion-wave fields to wave-like representations under arbitrary excitation conditions. By introducing a Fredholm integral formulation derived from the heat diffusion equation, the proposed method extends the virtual-wave concept beyond the conventional Dirac pulse assumption to encompass pulse, harmonic, and chirp-pulsed excitations. This generalization enables the use of flexible excitation schemes and bridges the gap between diffusion-dominated thermography and wave-based imaging methodologies. Theoretical analysis demonstrates that the transformation intrinsically preserves causality and thermodynamic irreversibility while enabling the reconstruction of propagative features from inherently diffusive signals. The ill-posed inverse problem is addressed through tailored regularization strategies, where ADMM is employed for sparse, time-localized responses induced by pulsed excitations, and TSVD is adopted for spectrally compact harmonic responses. This establishes a clear relationship between excitation design and inversion methodology, providing both physical insight and computational robustness. Numerical simulations confirm that the proposed approach successfully reconstructs virtual wave fields with clear wavefronts and reflection features, overcoming the fundamental limitations of conventional time- and frequency-domain thermography, such as blurring, lack of spatial resolution, and ambiguous depth interpretation. Experimental validation on CFRP samples with embedded delamination defects further demonstrates significant improvements in defect detectability, boundary sharpness, and depth localization. In particular, the transformation enables the recovery of reflection-like echoes and a linear depth interpretation, which are not accessible in classical diffusion-based frameworks. Compared with existing photothermal coherence tomography techniques, the proposed generalized virtual-wave reconstruction exhibits superior robustness and accuracy, especially under broadband excitations such as chirp-pulsed signals. The method not only enhances contrast-to-noise ratio but also eliminates spectral aliasing and nonlinear depth ambiguities, thereby enabling true tomographic reconstruction analogous to ultrasonic imaging. Overall, this work provides a physically grounded and mathematically consistent framework for converting diffusive thermal responses into wave-like signals, opening new possibilities for advanced signal processing, excitation design, and three-dimensional reconstruction in photothermal nondestructive evaluation. Future work may focus on extending the framework to heterogeneous and anisotropic materials, optimizing excitation waveforms, and integrating data-driven approaches to further improve reconstruction efficiency and robustness.

\begin{acknowledgments}
	This work was supported by the Adolf Martens Fellowship (Grant n. BAM-AMF-2025-1).
\end{acknowledgments}

\appendix
\section{Derivation of Virtual Wave Kernel}
By interchanging the order of integration, Eq.~\eqref{eq9} can be written as
\begin{equation}
	T(\mathbf{r},t)
	= \frac{c^2}{\kappa}
	\int_0^{\infty}
	T_{\mathrm{virt}}(\mathbf{r},t')
	\left[
	\frac{1}{2\pi}
	\int_{-\infty}^{\infty}
	e^{-c\sigma(\omega)t'}
	e^{i\omega t}\, d\omega
	\right] dt',
	\label{eq33}
\end{equation}
which naturally defines the kernel function
\begin{equation}
	K(t,t')
	= \frac{c^2}{\kappa}
	\frac{1}{2\pi}
	\int_{-\infty}^{\infty}
	e^{-c\sigma(\omega)t'} e^{i\omega t}\, d\omega.
	\label{eq34}
\end{equation}

Using $\sigma(\omega)=\sqrt{i\omega/\alpha}$, Eq.~\eqref{eq34} becomes
\begin{equation}
	K(t,t')
	= \frac{c^2}{\kappa}
	\frac{1}{2\pi}
	\int_{-\infty}^{\infty}
	\exp\!\left[-ct'\sqrt{\frac{i\omega}{\alpha}} + i\omega t\right] d\omega.
	\label{eq35}
\end{equation}

To evaluate the integral, we introduce the change of variable
$s=\sqrt{i\omega}$, such that $\omega=-is^2$ and $d\omega=-2is\,ds$. Under this transformation, the exponent can be rearranged by completing the square as
\begin{equation}
	-ct'\sqrt{\frac{i\omega}{\alpha}} + i\omega t
	= -t\left(s - \frac{ct'}{2\sqrt{\alpha t}}\right)^2
	- \frac{c^2 t'^2}{4\alpha t}.
	\label{eq36}
\end{equation}

The integral over $s$ then reduces to a standard Gaussian integral, yielding the factor
$t'/(2\sqrt{\pi}(\alpha t)^{3/2})$. Including prefactors and enforcing causality ($t>0$, $t'>0$), the kernel function is obtained as
\begin{equation}
	K(t,t')
	= \frac{c^2}{\kappa}
	\frac{t'}{2\sqrt{\pi}(\alpha t)^{3/2}}
	\exp\!\left(-\frac{c^2 t'^2}{4\alpha t}\right)
	H(t)H(t').
	\label{eq37}
\end{equation}

\begin{figure*}
	\centering
	\includegraphics[width=1.0\textwidth]{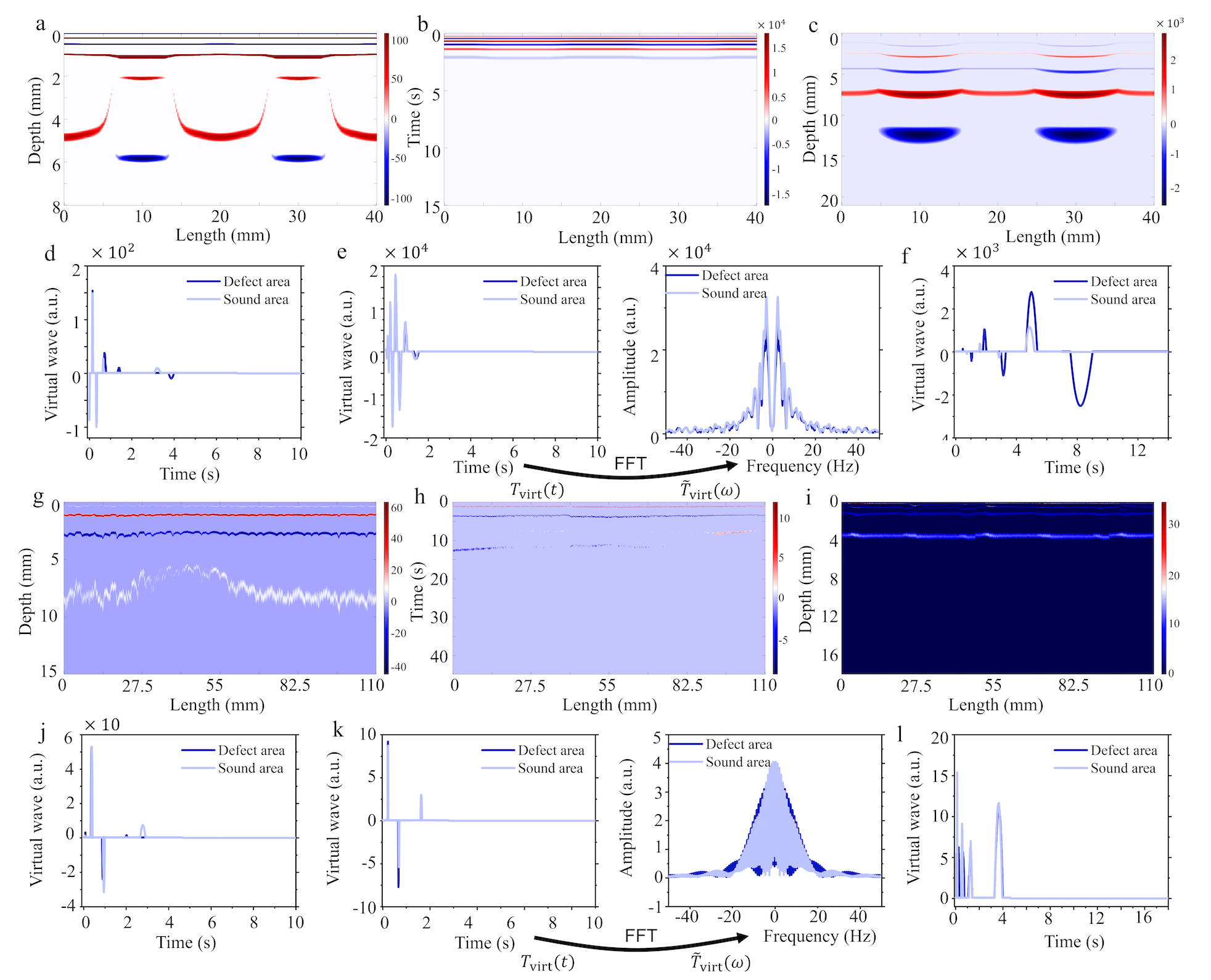}
	\caption{The special virtual wave transform results for simulation and photothermal experiments. a Virtual wave transform results of pulse excitation in simulation. b Virtual wave transform results of lock-in excitation in simulation. c Virtual wave transform results of chirp-pulsed excitation in simulation. d Virtual wave profiles vary with time of defect and sound areas in simulation under pulse excitation. e Virtual wave profiles vary with time and frequency of defect and sound areas in simulation under lock-in excitation. f Virtual wave profiles vary with time of defect and sound areas in simulation under chirp-pulsed excitation. g Virtual wave transform results of pulse excitation in experiments. h Virtual wave transform results of lock-in excitation in experiments. i Virtual wave transform results of chirp-pulsed excitation in experiments. j Virtual wave profiles vary with time of defect and sound areas in experiments under pulse excitation. k Virtual wave profiles vary with time and frequency of defect and sound areas in experiments under lock-in excitation. l Virtual wave profiles vary with time of defect and sound areas in experiments under chirp-pulsed excitation.}\label{fig7}
\end{figure*}

\section{Theory of Special Virtual Wave Transform}
Conventional virtual wave concept is based on the Dirac pulse excitation~\cite{ref32}. The thermal diffusion equation can be written as:
\begin{equation}
	\label{eq36}
	\left( \nabla^2 - \frac{1}{\alpha}\frac{\partial}{\partial t} \right) T(\mathbf{r},t)
	= -\frac{1}{\alpha} T_0(\mathbf{r}) \delta(t)
\end{equation}

where $\alpha = \frac{k}{\rho C}$ is the thermal diffusivity. The external incident heat is modeled as an initial condition instead of a boundary condition. The equivalence between these two boundary-value problems was validated by Zhu \textit{et al.}~\cite{ref33}. However, in practice, the external heat source cannot be a Dirac pulse. In general, the heating duration is on the order of milliseconds. Therefore, Eq.~\eqref{eq36} describes an idealized point heating source.

The wave equation describes the acoustic pressure $p(\mathbf{r},t)$ as a function of space $\mathbf{r}$ and time $t$:
\begin{equation}
	\label{eq37}
	\left( \nabla^2 - \frac{1}{c^2} \frac{\partial^2}{\partial t^2} \right) p(\mathbf{r},t)
	= -\frac{1}{c^2} \frac{\partial}{\partial t} \left[ p_0(\mathbf{r}) \delta(t) \right]
\end{equation}

where $c$ is the speed of sound, and $p_0(\mathbf{r})$ is the initial pressure distribution immediately after a short excitation pulse. A virtual wave field $T_{\mathrm{virt}}$ is defined using the same wave equation with initial temperature distribution $T_0(\mathbf{r})$ and an arbitrarily chosen $c$:
\begin{equation}
	\label{eq38}
	\left( \nabla^2 - \frac{1}{c^2} \frac{\partial^2}{\partial t^2} \right) T_{\mathrm{virt}}(\mathbf{r},t)
	= -\frac{1}{c^2} \frac{\partial}{\partial t} \left[ T_0(\mathbf{r}) \delta(t) \right]
\end{equation}

Both Eq.~\eqref{eq36} and Eq.~\eqref{eq38} can be transformed into the frequency domain:
\begin{equation}
	\label{eq39}
	\left( \nabla^2 - \sigma^2(\omega) \right) \theta(\mathbf{r},\omega)
	= -\frac{1}{\alpha} T_0(\mathbf{r})
\end{equation}

\begin{equation}
	\label{eq40}
	\left( \nabla^2 - \psi^2(\omega) \right) \theta_{\mathrm{virt}}(\mathbf{r},\omega)
	= -\frac{i\omega}{c^2} T_0(\mathbf{r})
\end{equation}
where
\begin{equation}
	\label{eq41}
	\sigma^2(\omega) \equiv \frac{i\omega}{\alpha}, 
	\qquad
	\psi(\omega) = \frac{\omega}{c}
\end{equation}
are the wavenumbers for the thermal diffusion-wave field and the virtual wave field, respectively. By substituting $\omega \rightarrow -i c \sigma(\omega)$ into Eq.~\eqref{eq40}, the relationship between the thermal diffusion-wave field and the virtual wave field in the frequency domain becomes:
\begin{equation}
	\label{eq42}
	\theta(\mathbf{r},\omega)
	= \frac{c}{\alpha \sigma(\omega)} \,
	\theta_{\mathrm{virt}}(\mathbf{r}, -i c \sigma(\omega))
\end{equation}

Therefore, the temperature in the time domain can be expressed as:
\begin{equation}
	\label{eq43}
	T(\mathbf{r},t)
	= \frac{1}{2\pi} \int_{-\infty}^{\infty}
	T_{\mathrm{virt}}(\mathbf{r}, t') \, K(t,t') \, dt'
\end{equation}
where the kernel function is given by:
\begin{equation}
	\label{eq44}
	K(t,t') \equiv \frac{c}{\sqrt{\pi \alpha t}} 
	\exp\left( -\frac{c^2 t'^2}{4 \alpha t} \right), \quad t > 0
\end{equation}

Equation~\eqref{eq43} establishes the relationship between the thermal diffusion-wave field and the virtual wave field in the time domain. This is a Fredholm integral equation of the first kind. By solving Eq.~\eqref{eq43} for $T_{\mathrm{virt}}$, the virtual wave field can be reconstructed.

The special virtual wave transform (SVWT) results are shown in Fig.~\ref{fig7}. SVWT exhibits excellent performance in the simulation under Dirac pulse excitation (Fig.~\ref{fig7}a). However, it completely failed in modulated excitations including lock-in and chirp-pulsed excitations (see Fig.~\ref{fig7}b-c and Fig.~\ref{fig7}e-f). Comparing with GVWR and SVWT results, it is obvious that GVWR results have higher defect contrast (see Fig.~\ref{fig7}d and Fig.~\ref{fig5}a). The same conclusion was demonstrated in experiments (see Fig.~\ref{fig7}g-l).

The failure of the special virtual wave transform (SVWT) can be traced back to its overly restrictive physical assumptions. In its original formulation, the SVWT is derived under the idealized Dirac pulse excitation, where the thermal field is treated as the Green’s response of the diffusion equation to an impulsive source. This assumption enables an elegant mapping between diffusion and wave equations, but it also fundamentally limits the applicability of the method. In practice, thermal excitation always has a finite temporal width and often involves complex boundary-driven heat flux rather than an initial-value formulation. As a consequence, the measured temperature field is not a simple convolution with a Dirac Green’s function, but rather a superposition of temporally distributed responses. The SVWT, which implicitly inverts a delta-driven kernel, becomes mismatched to the actual physics, leading to systematic reconstruction errors and loss of resolution. Moreover, the SVWT assumes a homogeneous and isotropic medium with constant thermal diffusivity. Any deviation from this assumption—such as layered structures, anisotropy, or spatially varying parameters—breaks the separability required for the transform. This results in model inconsistency, where the transformed field no longer corresponds to a physically meaningful wave analogue. The artifacts and instability observed in reconstructions are direct consequences of this mismatch. Another critical limitation lies in the ill-posed nature of diffusion inversion. The SVWT effectively attempts to reverse the diffusion operator by amplifying high-frequency components to mimic wave propagation. However, diffusion intrinsically suppresses high-frequency information, making such inversion exponentially unstable. As a result, the SVWT is highly sensitive to noise and measurement errors, particularly for deeper structures. In contrast, the generalized virtual wave transform (GVWT) succeeds because it relaxes these restrictive assumptions and reformulates the problem in a more physically consistent framework. Instead of relying on a Dirac excitation, the GVWT accommodates arbitrary temporal excitation by explicitly incorporating the source term or boundary conditions into the inversion process. This allows the transform to remain valid under realistic experimental conditions. Furthermore, the GVWT can be interpreted as an operator-based mapping that preserves the full structure of the diffusion process, rather than enforcing a simplified analytical correspondence. By working with generalized Green’s functions or integral representations, it naturally extends to heterogeneous and anisotropic media, thereby reducing model mismatch. Importantly, the GVWT does not attempt to fully invert the diffusion operator in a naive sense. Instead, it implicitly introduces regularization through its formulation, balancing resolution enhancement with stability. This leads to improved robustness against noise and better reconstruction fidelity, especially for deeper or low-contrast features. In summary, the contrast between SVWT and GVWT reflects a broader principle: methods based on idealized analytical transforms tend to fail when confronted with realistic physical complexity, whereas operator-consistent and model-aware formulations provide a more reliable pathway for inverse diffusion problems.

\end{document}